\documentclass{PoS}

\title{INTEGRAL view of the Extragalactic Sky}

\ShortTitle{INTEGRAL view of the Extragalactic Sky}

\author{\speaker{A. Malizia}\\
                   INAF IASF--Bologna\\
        E-mail: \email{malizia@iasfbo.inaf.it}}

\author{L.  Bassani\\
          INAF IASF--Bologna\\
          E-mail: \email{bassani@iasfbo.inaf.it}}
          
\author{R. Landi\\ 
INAF IASF--Bologna\\
Email: \email{landi@iasfbo.inaf.it} }       

\author{A. Bazzano\\
INAF IASF-Roma\\
Email: \email{angela.bazzano@iasf-roma.inaf.it}}

\author{A. J. Bird\\
University of Southampton\\
Email: \email{A.J.Bird@soton.ac.uk}}

\author{F. Panessa\\
INAF IASF-Roma\\
Email: \email{francesca.panessa@iasf-roma.inaf.it}}   

\abstract{Presented in this work is the most comprehensive INTEGRAL AGN sample: it includes 272 active galaxies for which optical identifications and redshifts are 
available.  2-10 keV fluxes and  column density measurements are also collected for all the sample sources. For 33 new hard X-ray discovered AGN with no previous X-ray coverage, 
XRT and XMM data analysis is reported for the first time in this work. Examples of future studies which are being developed using  this large X-ray selected sample 
are presented together with some early results. }

\FullConference{The Extreme and Variable High Energy Sky - extremesky2011,\\
		September 19-23, 2011\\
		Chia Laguna (Cagliari), Italy}

\begin{document}

\section{The INTEGRAL AGN sample}
From 723 sources listed in the 4th INTEGRAL/IBIS survey [1] we took a census of 234 AGN considering only the extragalactic objects for which
a firm optical identification and redshift measurement  is available.
In order to a have a complete view of INTEGRAL extragalactic sky we have also cross-correlated the 4th IBIS catalogue by Bird et al. 2009 with the 
INTEGRAL all-sky survey by Krivonos et al. [4] updated on the web site, and among their 211 objects identified as AGN, we found 38 extra sources not included in our list.
With the addition of these extra objects the final INTEGRAL AGN catalogue presented in this work lists 272 galaxies (last update March 2011). 
Note that new sources are continuously coming  in [6].
In figure 1 (left) the whole sample is reported in the classical 20-100 keV luminosity $vs$ redshift plot.
This sample represents the most comprehensive INTEGRAL AGN catalogue and this has two strengths. First: all sources, except for one BL LAC
object, have optical spectra which means  secure identifications and measured redshifts; and second: all 
sources (except for one case) have X-ray data and so 2-10 keV fluxes and N$_H$ estimates available.

The 272 AGN belong to the following optical classes: 
132 objects are Seyfert of type 1 (Seyfert 1, 1.2, 1.5 and NLS1) , 116 are Seyfert of type 2 (Seyfert 1.9-2,  and XBONG) and 22 are blazars (BL Lac and QSO).

In order to assign to each object the most appropriate optical class, a cross check between the class reported in the NED archive and that reported in the last edition of the 
Veron-Cetty, Veron  extragalactic catalogue [9] has been performed with the exception of those objects for which the optical classification has been provided by our own
identification works [6, and reference therein]. In case of different classification the more recent one has been preferred and for subclass
assignments (Sy1.2, Sy1.5, Sy1.8, Sy1.9) the diagnostic criteria of Winkler  [10] have been adopted.

\begin{figure}
\includegraphics[width=0.5\linewidth]{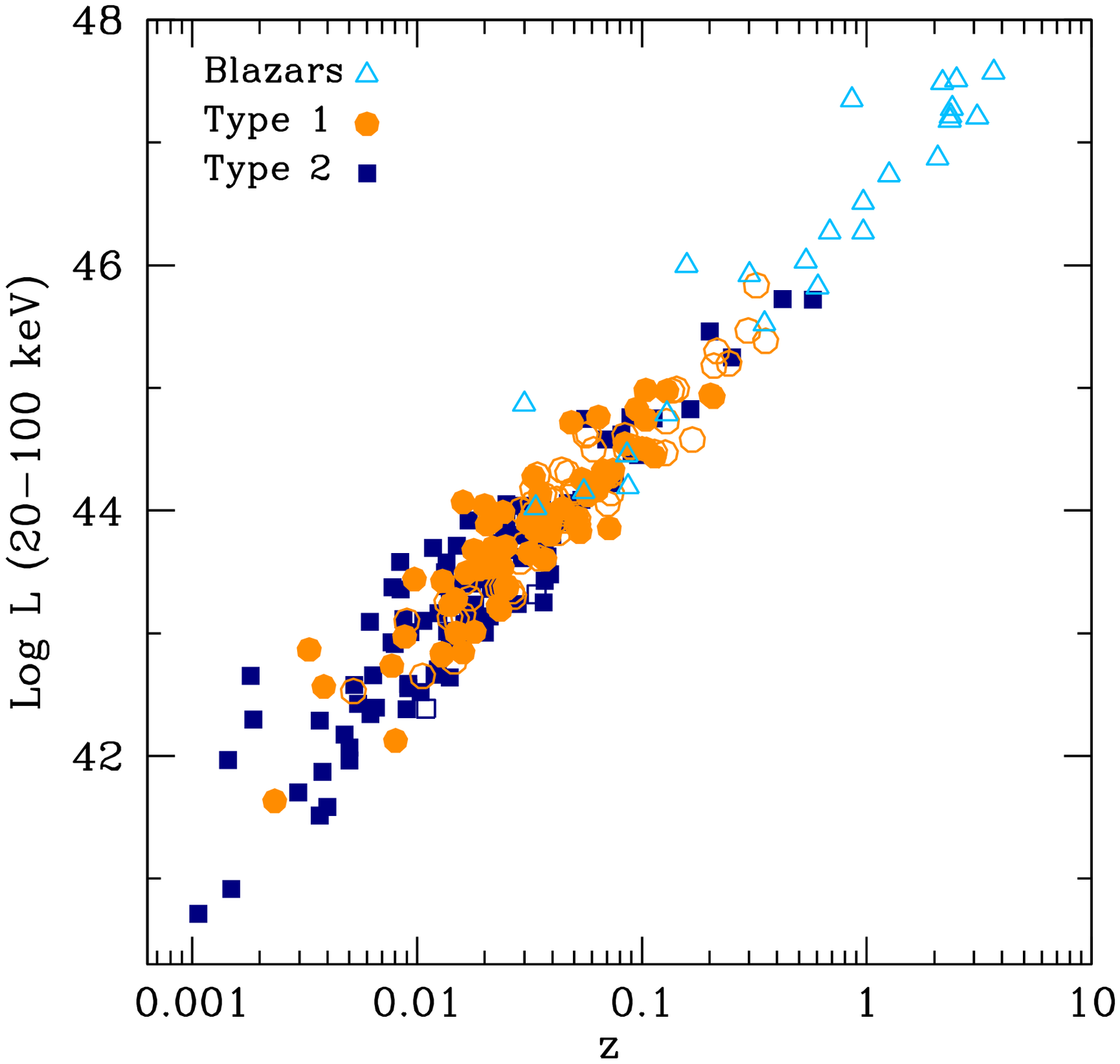}
\includegraphics[width=0.5\linewidth]{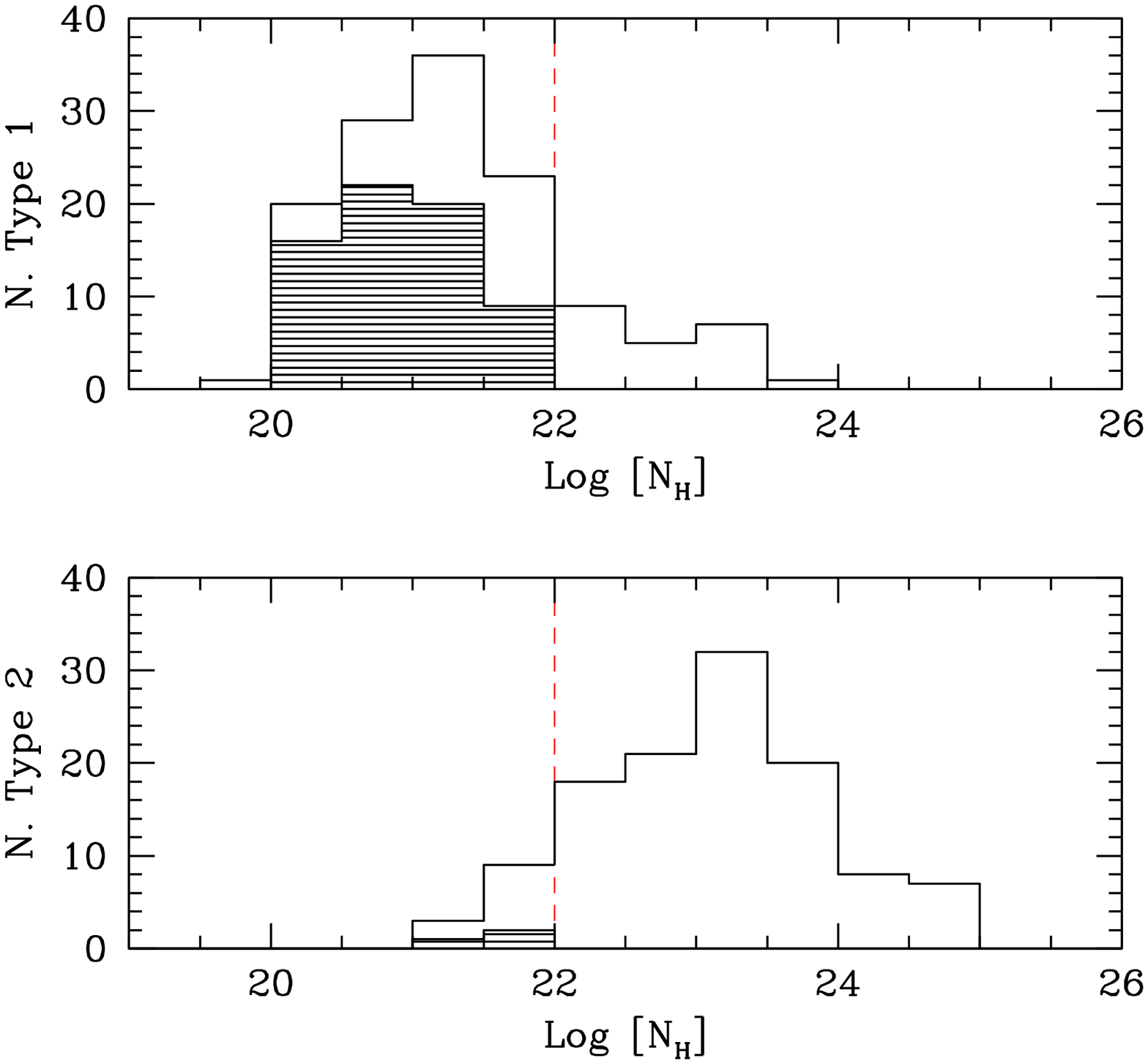}
\caption{\small
{{\it Left panel}: Hard X-ray luminosity vs redshift for all the INTEGRAL AGN sample. Circles are type 1 objects, squares are type 2 and triangles are blazars. 
Open symbols are objects where no intrinsic absorption have been measured.
{\it Right panel}: Column density distributions in type 1 (up) and type 2 (down). Dashed bins represent upper limit measurements including Galactic values.}}
\label{fig1}
\end{figure}

\section{X-ray data analysis and results}
Among the 272 AGN there are 33 sources which are new hard X-ray emitters (discovered by INTEGRAL/Swift) and therefore they had no X-ray (2-10 keV) coverage
prior to this work.
For these sources we analysed X-ray data available from Swift  and XMM-Newton observations.

Swift-XRT data are available for 31 sources and they have been analysed in this work  for the first time. 
XMM data of RX J0137.7+5814 and IGR J13415$+$3033 are also considered as, for the first source, no X-ray published data have been found,
and for the second, a re-analysis has been performed getting quite different results form those obtained by [7] using XRT data alone.

XRT data reduction was accomplished using the XRTDAS standard data pipeline package (xrtpipeline v. 0.12.6), 
in order to produce screened event files.
Furthermore, to univocally identify the X-ray counterpart, an accurate imaging analysis has been performed for those new AGN first detected at high energies 
($>$20 keV) by INTEGRAL/IBIS. For each observation we analysed the 0.3-10 keV image to search for sources detected 
at a confidence level $> 3\sigma$ both within the 90\% and 99\% IBIS error circles. 
If more sources were present within these IBIS uncertainties, we restricted the image analysis to the 3-10 keV band to spot those sources with the hardest spectra 
(i.e. those with detection above 3 keV), since these are most likely to be the counterparts of IBIS objects. 
Then, we estimated the X-ray positions using the task xrtcentroid, providing in this way for each source the position  at arsec level accuracy in order to be sure of the optical identification.
Events for spectral analysis were extracted within a circular region of radius 20$^{\prime\prime}$, centered on the source position, which encloses about 90\% of the PSF at 1.5 keV, while the background was taken from various source-free regions close to the X-ray source of interest.
In all cases, the spectra were extracted from the corresponding event files using the XSELECT software and binned using grppha in an appropriate way, so that the $\chi^{2}$ statistic could be applied. 
We used version v.011 of the response matrices and create individual ancillary response files arf using xrtmkarf v. 0.5.8. 
The energy band used for the spectral analysis, performed with XSPEC v. 12.5, depends on the statistical quality of the data and typically ranges from 0.3 to 8 keV.

For two sources of our sample we use X-ray data acquired with XMM-Newton and they were reprocessed using the Standard Analysis Software (SAS) version 9.0.0 
employing the latest available calibration files. 
Only patterns corresponding to single and double events (PATTERN$\leq$4) were taken into account and the standard selection filter FLAG=0 was applied. 
For each observation, we analysed the XMM-Newton EPIC (MOS plus pn) images to search for X-ray sources which fall inside the IBIS error box and are therefore likely IBIS counterparts. 
Next we obtained X-ray spectra in the 0.5-12 keV band. Source counts were extracted from circular regions of radius 20$^{\prime\prime}$ centered on the source
while background spectra were extracted from circular source-free regions of 80$^{\prime\prime}$ radius. 
The ancillary response matrices (ARFs) and the detector response matrices (RMFs) were generated using the XMM SAS tasks arfgen and rmfgen while the spectral channels 
were rebinned in order to achieve a minimum of 20 counts in each bin.

In order to estimate the X-ray spectral parameters of these newly discovered sources, we fit the X-ray spectrum  using a baseline model composed by a primary continuum modeled with a 
power law absorbed by the Galactic column density calculated through XSPEC [2].
To measure the amount of cold absorption intrinsic to the source, an extra column has been added when required.
In a few cases, due to the poor statistical quality of the data, the photon index was fixed to $\Gamma$ = 1.8 which represents a canonical AGN value 
so as to allow a measurement of the intrinsic column density.
In table 1 the data analysis results are reported together with the optical classification of each object; the last two are those for which XMM data have been reduced.

The set of AGN analysed for the first time here, are well representative of the entire INTEGRAL sample.
A few peculiar objects  emerge from table 1, like IGR J0145+6437 which is optically classified as Sy2 but shows no intrinsic absorption, and IGR J03249+4041 which is a pair of galaxies both at $\simeq$0.047, both 
classified as Sy 2 and both emitting at hard X-ray energies.

\begin{table*}
\begin{center}
\centerline{Results of the spectral analysis of the XRT/XMM data }
\small
\begin{tabular}{llcccc}
\hline
\hline
Source                              & Class           &  $N_{H(intr)}$                         & $\Gamma$                        &  $F_{(2-10~ keV)}$              &  $\chi^2/\nu$  \\
                                         &                     &  ($10^{22}$ cm$^{-2}$)        &                                           &  ($10^{-11}$ erg cm$^{-2}$ s$^{-1}$)  &                \\
\hline
\hline
IGR J00158+5605          & Sy1.5                        &      --                                                                 & $2.30\pm0.34$                         &  0.81          &  3.7/6           \\
IGR J01545+6437          & Sy2                           &     --                                               & [1.8]                                             &  0.0097     &  3.8/3           \\
RBS 345                           & Sy1                         &     --                                               & $1.62^{+0.24}_{-0.22}$           &  0.26         &  17.2/8          \\  
IGR J03249+4041-SW$^{(\dagger)}$   & Sy2    &        $>$ 0.15                                 & $ 1.53^{+0.47}_{-0.39}$          &  0.12         &   7.56/9       \\
IGR J03249+4041-NE$^{(\dagger)}$   & Sy2     &   $3.04^{+1.54}_{-0.99}$               &   [1.8]                                            & 0.09         &   4.35/6    \\
IGR J03334+3718          & Sy1.5                          &   $0.14^{+0.06}_{-0.06}$                &  1.86$^{+0.18}_{-0.17}$          &   0.61       &  49.1/35  \\
LEDA 15023                    & Sy2                         &  $30.1^{+9.3}_{-9.1}$                     & $2.21^{+0.64}_{-0.50}$          &  0.13        &  8.1/8          \\
LEDA 075258                 & Sy1                          &  0.0057$^{+0.002}_{-0.002}$         &  1.77$^{+0.09}_{-0.09}$          &   0.55       & 109.8/109   \\
IGR J06058-2755$^{(a)}$   & Sy1.5                  &                    --                                      &  1.71$^{+0.09}_{-0.09}$          &  0.66       & 39.0.50          \\
IGR J08557+6420$^{(b)}$  & likely Sy2             &    19.6$^{+4.5}_{-4.0}$                    &  [1.8]                                            &  0.28        &  6.3/8            \\
IGR J08558+0814          & Sy1                          &     --                                                   &  $2.24\pm0.46$                         &  0.0065    &  4.6/13          \\
IGR J09253+6929          & Sy1.5                       & 14.8$^{+28.0}_{-11.0}$                & [1.8]                                              & 0.045      & 0.22/2 \\
IGR J09446--2636         & Sy1.5                       &    --                                                   & $1.66\pm0.09$                           &  0.50       &  38.8/32         \\
IGRJ13187+0322           &  QSO                       &        --                                            & [1.8]                                                & 0.015     &  3.0/6           \\
IGRJ14175--4641          & Sy2                          &   $75.5^{+283.0}_{-40.0}$          & [1.8]                                                &  0.095     &  3.0/5            \\
IGR J14301--4158         & Sy2                          &  $1.4^{+0.8}_{-0.5}$                   & $1.83^{+0.50}_{-0.58}$              &  2.70        &  5.5/11           \\
IGR J15311--3737         & Sy1                          &    --                                                 & $1.73\pm0.16$                              &  1.86       &  8.4/11            \\
IGR J15549--3740         & Sy2                          &  $5.1^{+2.0}_{-1.3}$                    & [1.8]                                                 &  0.29       &  8.5/9              \\
IGR J16426+6536          & NLSy 1                    &    --                                                  & [1.8]                                                &  0.0085   &  4.3/6             \\
SWIFT J1650.5+0403    & Sy2                          &  $4.8^{+1.4}_{-1.3}$                   & $1.87^{+0.21}_{-0.37}$             & 0.32         &  23.6/32         \\
IGR J17036$+$3734      & Sy1                         &  --                                               & $2.30^{+0.32}_{-0.21}$             & 0.21        & 4.2/5 \\
IGR J17476$-$2253         & Sy1                       &    --                                                  & $1.91^{+1.00}_{-0.91}$            & 0.26          &  2.7/3            \\
1RXS J175252.0$-$053210 & Sy1.2                &   --                                               & $1.93^{+1.16}_{-1.00}$                & 0.26         & 2.3/2  \\
IGR J18311--3337         & Sy2                         &  $1.4^{+5.0}_{-1.0}$                   & [1.8]                                               & 0.026        &  3.2/6            \\ 
IGR J19077--3925         & Sy1.9                      &  $0.14^{+0.10}_{-0.08}$            & $1.78^{+0.24}_{-0.22}$             & 0.36           &  11.4/18           \\ 
IGR J19118--1707         & LINER                     &  $10.4^{+8.1}_{-4.9}$                 & [1.8]                                               & 0.14           &  2.1/4            \\
PKS 1916--300              & Sy1.5/1.8                &    --                                                 & $2.04\pm0.07$                           & 0.62           &  57.7/66          \\
1RXS J192450.8--291437   & BL Lac              &  $0.073\pm0.03$                        & $1.95\pm0.10$                           & 0.81           &  104.1/89        \\
QSO B1933--400           & QSO                       &  $0.13^{+0.13}_{-0.10}$            & $1.84^{+0.19}_{-0.31}$            & 0.17           &  11.7/13          \\
IGR J19491-1035          & Sy1.2                      &  $1.80^{+0.79}_ {-0.73}$         &  $1.79^{+0.41}_{-0.42}$           & 0.56            & 19.68/14\\
SWIFT J2044.0+2832   & Sy1                          &  --                                                   & 1.94$^{+0.05}_{-0.05}$            &  0.71         & 109.1/115  \\
1RXS J211928.4+33359  & Sy1.5                    &  $0.46^{+0.08}_{-0.07}$            & $1.78^{+0.18}_{-0.17}$           & 0.59           &  80.2/61            \\
\hline
\hline
RX J0137.7+5814           & BL Lac                  &               --               &    2.12$^{+0.04}_{-0.04} $    &            1.01                                                &     224/210   \\
IGR J13415$+$3033       &  Sy 2                     &     29.70$^{+3.51}_{-2.00}$       &     1.97$^{+0.16}_{-0.18}$    &             0.25                                             &    426/397  \\
\hline
\hline
\end{tabular}
\begin{list}{}{}
$^{\dagger}$ interacting galaxies, here a tentative spectral analysis has been performed \\
$^{a}$ Best-fit model includes an absorption edge at $\sim$0.52 keV\\
$^{b}$ Best-fit model includes a second power law of $\Gamma$= 1.8 (fixed) to account for the excess observed below 1 keV. \\
\end{list}
\end{center}
\end{table*}

\section{INTEGRAL AGN sample: Exploitation Studies}
\subsection{X-ray absorption versus Optical classification}
The unified model for Seyfert galaxies explains their different observed characteristics with their orientation with respect to the observer line of sight.
An AGN is classified as type 1  if there is the presence of broad permitted lines in its optical spectrum while it is classified as type 2 if it does not show these broad components. 
In the type 2 Seyfert the broad line region is hidden by absorbing material which should largely correspond to that measured in the X-ray band. Therefore heavy absorption is expected in type 2 and less or even none in type 1. 
It is well known that there is a significant number of AGN for which the expected optical and X-ray characteristics do not match and we know cases of AGN of type 1 where strong absorption is measured and,
on the other hand, there are also AGN showing in their optical spectra only narrow emission lines 
but displaying only mild or even absent absorption in their X-ray spectra.

The INTEGRAL sample of AGN  is an ideal tool  to investigate this issue since the hard X-ray selection is almost unbiased with respect  to the absorption.
To this end we have performed the column density distribution dividing our sample sources in type 1 (including also Sy1.2-Sy1.5) and type 2 (including also Sy1.9).
The histograms of N$_H$ in type 1 and type 2 Seyfert galaxies in the INTEGRAL AGN sample is shown in figure 1 (right).
If we assume the dividing line between non-absorbed and absorbed AGN lies at N$_H$ = 10$^{22}$  cm$^{-2}$ [\footnote{it corresponds to the column density needed to hide the BLR for clouds having a standard gas-to-dust ratio [8]}],
it is clearly visible from figure 1 (right) that  a number (23) of type 1 objects show intrinsic absorption higher than 10$^{22}$ cm$^{-2}$, and a number (11) of type 2 
have column density lower than 10$^{22}$ cm$^{-2}$ or even none in excess to the Galactic one (dashed bins in figure). If we take into account only the 'pure' Seyfert 
galaxies (i.e. excluding LINERS, XBONG), the total number of AGN which do not fit into the unified scheme  are 33/250 which correspond to 13\% of the sample and this 
percentage is quite close to that typically found in the soft X-ray selected samples of AGN [3].

\subsection{X-ray absorption versus host galaxy axial ratio}
It is becoming increasingly clear that the study of AGN is strongly connected with the properties of the host galaxies and, in particular, the inclination of the host galaxy
can play a role in relation to the X-ray absorption. The inclination is determined by the axial ratio which  is defined as the semi-minor (b) over semi-major axis (a) so that b/a=1 corresponds to a face-on galaxy and  b/a$\simeq$0  to an edge-on galaxy.
We have collected from the literature all the available axial ratios for the entire INTEGRAL AGN sample and
we have found a different behavior between type 1 and type 2 objects, with only type 1 AGN showing a deficit at low values of b/a. 
This has been investigated in terms of absorption suggesting the existence of an extra absorbing component on a larger scale than the torus 
assumed in the unified theory.

\subsection{X-ray absorption versus galactic structures and interactions}
Another possible connection between the X-ray absorption and the large scale structures could be represented by the presence of
large scale gravitational torques such as in barred and interacting galaxies as they can transport gas close to the center of the AGN 
and this gas concentration can play a role in the X-ray obscuration [5]. 
Using our sample we do not find any correlation between absorption and the presence of bars and/or interactions, despite the fact that
$\sim$40\% of our galaxies are strongly barred and 38\%  interacting.

These are some of the results that can be obtained with our large sample,
 highlighting the potential of statistical and population studies of hard X-ray selected AGN.
 Further works are in progress.\\

{\bf Acknowledgments}\\
The authors acknowledge support by INTEGRAL ASI I/033/10/0 and ASI/INAF I/009/10/0


\begin{thebibliography}{99}
\bibitem{}Bird A. J., Bazzano A., Bassani L., et al. 2010, ApJS, 170, 175 
\bibitem{}Dickey, J. M., \& Lockman, F. J. 1990, IRA\&A, 28, 215
\bibitem{}Garcet O., Gandhi P., Gosset E., et al. 2007, A\&A, 474, 473
\bibitem{}Krivonos R., Revnivtsev M., Tsygankov S., et al. 2007, A\&A, 475, 775 
\bibitem{}Maiolino R., Risaliti G, Salvati M. 1999, A\&A, 341, L35
\bibitem{}Masetti N., Parisi P., et al. 2011, A\&A submitted
\bibitem{}Rodriguez J., Tomsick J. A.,  Bodaghee A. 2010, A\&A, 517, 14
\bibitem{}Silverman J. D., Green P. J., Barkhouse W. A., et al. 2005, ApJ, 618, 123
\bibitem{}Veron-Cetty, M.-P.; Veron, P. 2010, A\&A, 518, 10
\bibitem{}Winkler H. 1992, MNRAS, 267, 677


\end{thebibliography}
\end{document}